\documentclass[prd,twocolumn,showpacs,preprintnumbers,superscriptaddress,floatfix,nofootinbib]{revtex4-1}
\usepackage{amsfonts}
\usepackage{graphicx,,booktabs}
\usepackage{amssymb}
\usepackage{amsmath,txfonts}
\usepackage{graphicx}
\usepackage{dcolumn}
\usepackage{color}
\usepackage{bm}
\usepackage[colorlinks,
            citecolor=blue,
            anchorcolor=green,
            menucolor=orange,
            linkcolor=red,
            filecolor=red,
            runcolor=pink,
            urlcolor=blue,
            frenchlinks=red]{hyperref}

\allowdisplaybreaks[4]

\begin{document}
\title{Systematic study on  production of hidden-bottom
pentaquark via $\gamma p$ and $\pi ^{-}p$ scatterings}
\author{Xiao-Yun Wang}
\thanks{xywang01@outlook.com}
\affiliation{Department of physics, Lanzhou University of Technology,
Lanzhou 730050, China}
\author{Jun He}
\thanks{Corresponding author: junhe@njnu.edu.cn}
\affiliation{Department of Physics and Institute of Theoretical Physics, Nanjing Normal University,
Nanjing, Jiangsu 210097, China}
\author{Xurong Chen}
\thanks{xchen@impcas.ac.cn}
\affiliation{Institute of Modern Physics, Chinese Academy of Sciences, Lanzhou 730000,
China}
\affiliation{University of Chinese Academy of Sciences, Beijing 100049}
\affiliation{Institute of Quantum Matter, South China Normal University, Guangzhou 510006, China}

\begin{abstract}
In this work, the production of hidden-bottom pentaquark $P_{b}$ states via $\gamma p$ and $%
\pi ^{-}p$ scatterings is studied within an effective Lagrangian approach and
the vector-meson-dominance mechnism. For the $P_{b}$  production in  the process $\gamma p\rightarrow
\Upsilon p$, the dipole Pomeron model is employed to calculate the
background contribution, and the
experimental data can be well described. For the process $\pi ^{-}p\rightarrow \Upsilon n$, the
Reggeized $t$-channel with $\pi $ exchange is considered as the main
background for the $P_{b}$  production. The cross section from $t$%
-channel $\pi $ exchange is very small due to  weak coupling  of
$P_{b}$ state to the $\pi \pi $ channel predicted theoretically. Near the threshold, two-peak
structure from the states $P_{b}(11080)$ and $P_{b}(11125)$ can be observed if
energy bin width is chosen as 0.01 GeV, and the same result is obtained in the $\pi ^{-}p
$ scattering. Moreover, by taking the branching ratio of Br$[{P_{b}\rightarrow \pi N}]\simeq 0.05\%$, the numerical result shows that
the average value of  cross section from the $P_{b}(11080)$ state
produced in the $\gamma p$ or $\pi ^{-}p$ scattering reaches at least 0.1 nb
with a bin of 0.1 GeV. Even if we reduce the branching ratio of the $P_{b}$ state into $\pi N$ channel by one order, the theoretical average of the cross section from $%
P_{b}(11080)$ production in $\pi ^{-}p$ scattering can reach the order of
0.01 nb with a bin of 0.1 GeV, which means that the signal can still
be clearly distinguished from the background. The experimental measurements
and studies on the hidden-bottom pentaquark $P_{b}$ state production in the $\gamma p
$ or $\pi ^{-}p$ scattering near-threshold energy region around $W\simeq 11$
GeV are strongly suggested, which are accessible at COMPASS and JPARC. Particularly, the
result of the photoproduction suggests that it is very promising to observe the hidden-bottom pentaquark at proposed EicC facility in China.
\end{abstract}

\maketitle

\section{Introduction}

Recently, three narrow hidden-charm pentaquark states, $%
P_{c}(4312),P_{c}(4440)$ and $P_{c}(4457)$, were observed by the LHCb
Collaboration~\cite{Aaij:2019vzc}. It is interesting that $P_{c}(4440)$ and $%
P_{c}(4457)$ are two peaks, which are revolved from the previously discovered $P_{c}(4450)$~\cite{Aaij:2015tga}. Moreover, it is found that these three $P_{c}$
states are quite narrow, and can be clearly seen in the $J/\psi p$ invariant
mass spectrum~\cite{Aaij:2019vzc}. As it was mentioned in Ref.~\cite%
{Aaij:2019vzc} that these three $P_{c}$ states are very good candidates of
molecular states, but the explanations of them as tightly bound pentaquarks cannot be ruled out. Soon after these $P_{c}$ states were discovered, the attempts to understand their internal structure were performed  in a large amount of  theoretical
models~\cite%
{Liu:2019tjn,He:2019ify,Chen:2019asm,Chen:2019bip,Huang:2019jlf,Ali:2019npk,Xiao:2019mst,Guo:2019kdc,Xiao:2019aya,Guo:2019fdo,Fernandez-Ramirez:2019koa,Zhu:2019iwm,Wu:2019rog,Voloshin:2019aut,Lin:2019qiv,He:2019rva,Yuan:2012wz}%
. However, more experimental and theoretical research is still needed to justify  which one of these explanations  is the real origin of  these exotic states.  Moreover, as shown in the Review of  Particle Physics (PDG)~\cite{Tanabashi:2018oca}, although many
exotic hadrons were listed, most of them were observed in the electron-positron  or proton-proton collision.  Hence, production of  these exotic hadrons in more different mechanisms is very important to  understand their nature.

Up to now, the hidden-charm pentaquarks were only observed in the $\Lambda
_{b}$ decay at LHCb. Production of the pentaquarks in other ways is very
helpful to obtain the definite evidence for the $P_c$ states as genuine states.
Different methods were proposed to detect the newly observed  $P_{c}$ states
 including photoproduction~\cite{Huang:2013mua,Wang:2019krd} and $\pi ^{-}p$
scattering~\cite{Wang:2019dsi}. Recently, the measurement of the process $\gamma p\rightarrow J/\psi
p$ was reported by the GlueX Collaboration \cite{Ali:2019lzf}. Although the signal of the $P_{c}$ state was not detected due to insufficient experimental precision, the branching ratio of the $P_{c}$ state decaying to $J/\psi p$ given by
the experiment is consistent with our previous theoretical prediction~\cite{Wang:2019krd}.

The hidden-charm pentaquark is a quark system composed of three light quarks and a charm quark pair. It is natural to extend it to the bottom sector to predict  hidden-bottom pentaquark which contains a bottom quark pair.  Such extension is more reasonable in the molecular state picture which is widely adopted to explain the $P_c$ states~\cite{Chen:2019asm,Liu:2019tjn,He:2019ify,Huang:2019jlf,Xiao:2019aya,Guo:2019fdo,Voloshin:2019aut,He:2015cea,Yang:2011wz}. Due to the existence of the heavy quark symmetry, the  $\Sigma^{(*)}_bB$ interaction is analogous to the $\Sigma^{(*)}_c\bar{D}$ interaction, which will lead to the existence of  hidden-bottom pentaquark.  Such similarity has been found in the comparison between the states $Z_b(10610)$ and $Z_b(10650)$ in bottom sector and the states $Z_c(3900)$ and $Z_c(4020)$ in charm sector. Based on such justification,  many
predictions about the hidden-bottom pentquarks have been given within different theoretical models~\cite{Wu:2010rv,Xiao:2013jla,Wu:2017weo,Yamaguchi:2017zmn,Anwar:2018bpu,Yang:2018oqd,Huang:2018wed,Wang:2019ato,Gutsche:2019mkg}.

It is interesting to study the possibility to observe such hidden-bottom pentaquark in future experiment. The $P_c$ states were observed in the $\Lambda_b$ decay at LHCb. However, it is impossible to find an analogous decay for hidden-bottom pentaquark.  If we recall that such pentaquarks are composed of three light quarks and a heavy quark pair, it is reasonable to produce them by exciting the nucleon by photon or pion meson to drag out a hidden-bottom quark pair.   Hence, in the current work, we  will study the possibility of  production of hidden bottom pentaquark $P_{b}$ states via $\gamma p$ and $%
\pi ^{-}p$ scatterings.

Because such states are still not observed in experiment, we need the theoretical decay width to make the prediction. One notices that in a recent paper~\cite{Gutsche:2019mkg}, the authors
predicted a series of hidden-bottom pentaquark states based on the observed
three $P_{c}$ states within the hadron molecular state model, and the decay widths of these states into the $%
\Upsilon p$ channel were also provided in their work. The mass and decay properties of
the $P_{b}(11080)$ state predicted in Ref.~\cite{Gutsche:2019mkg} are  also consistent with the prediction about a  $N^{\ast }(11100)$ state in Ref.~\cite{Wu:2010rv}.

\begin{figure}[tbph]
\begin{center}
\includegraphics[scale=0.52]{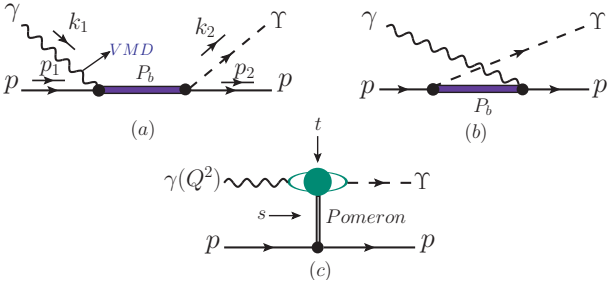}
\end{center}
\caption{Feynman diagrams for the reaction $\protect\gamma %
p\rightarrow \Upsilon p$.}
\label{Fig: Feynman1}
\end{figure}

In the current work,  the production of the hidden-bottom pentaquark $P_{b}$ states via
reactions $\gamma p\rightarrow \Upsilon p$ and $\pi ^{-}p\rightarrow \Upsilon n$
 will be investigated within the framework of an effective Lagrangian approach and the vector-meson-dominance (VMD) model.
The basic Feynman diagrams of the reaction $\gamma p\rightarrow \Upsilon p$
are illustrated in Fig.~\ref{Fig: Feynman1} where the pentaquarks $P_{b}$
are produced through $s$- and $u$-channels. Fig.~\ref{Fig:
Feynman1}(c) depicts the Pomeron exchange process, which is considered as
the mainly background contribution.
Considering the off-shell effect of the intermediate $P_{b}$ states, the $u$%
-channel contribution in $\gamma p$ or $\pi ^{-}p$ scattering will be
neglected in our calculation.

\begin{figure}[tbph]
\begin{center}
\includegraphics[scale=0.52]{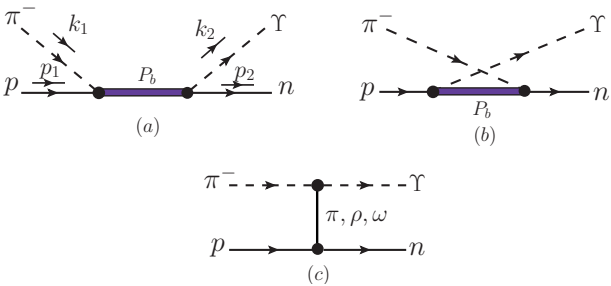}
\end{center}
\caption{Feynman diagrams for the reaction $\protect\pi %
^{-}p\rightarrow \Upsilon n$.}
\label{Fig: Feynman2}
\end{figure}

In Fig.~\ref{Fig: Feynman2}(a)-(b), the Feynman diagrams are presented for
describing the $P_{b}$ production in the reaction $\pi ^{-}p\rightarrow \Upsilon n$
 at  tree level. Moreover, the Reggeized $t$ channel
will be responsible for describing the main background, as depicted
in Fig. \ref{Fig: Feynman2}(c).

This paper is organized as follows. After the introduction, we present the
formalism including Lagrangians and amplitudes for the reactions $\gamma p\rightarrow
\Upsilon p$ and $\protect\pi ^{-}p\rightarrow \Upsilon n$  in Section II. The numerical results of the cross
section follow in Section III. Finally, the paper{\ ends} with a brief
summary.

\section{Formalism}

\subsection{Lagrangians for the $P_{b}$ production}

Since the $P_{b}$ states have not been found experimentally, in this work, the
spin-parity quantum number of the $P_{b}$ state is taken as $J^{P}=1/2^{-}$
as suggested in Refs.~\cite{Gutsche:2019mkg,Wu:2010rv} %.
. For describing the $P_{b}$ production in photon and pion induced processes,
the following Lagrangians are needed \cite%
{Wan:2015gsl,Wang:2015hfm,Kim:2011rm,Wang:2019krd,Wang:2019dsi},
\begin{eqnarray}
\mathcal{L}_{\gamma NP_{b}} &=&\frac{eh}{2m_{N}}\bar{N}\sigma _{\mu \nu
}\partial ^{\nu }A^{\mu }P_{b}+\mathrm{H.c.}, \\
\mathcal{L}_{P_{b}\Upsilon N} &=&g_{P_{b}\Upsilon N}\bar{N}\gamma _{5}\gamma
_{\mu }P_{b}\Upsilon ^{\mu }+\mathrm{H.c.}, \\
\mathcal{L}_{\pi NP_{b}} &=&g_{_{\pi NP_{b}}}\bar{N}\vec{\tau}\cdot \vec{\pi}%
P_{b}+\mathrm{H.c.},
\end{eqnarray}%
where ${N}$, $A$, $P_{b}$, $\pi $ and $\Upsilon $ are the nucleon, photon, $%
P_{b}$ state, pion, and $\Upsilon $ meson fields, respectively. And ${\tau }$
is the Pauli matrix.

The values of coupling constants $g_{P_{b}\Upsilon N}$ and $g_{_{\pi NP_{b}}}$ can be derived
from the corresponding decay width%
\begin{eqnarray}
\Gamma _{P_{b}\rightarrow \Upsilon N} &=&\frac{|\vec{p}_{\Upsilon }^{~%
\mathrm{c.m.}}|}{16\pi m_{P_{b}}^{2}}\left\vert \mathcal{M}%
_{P_{b}\rightarrow \Upsilon N}\right\vert ^{2}, \\
\Gamma _{P_{b}\rightarrow \pi N} &=&\frac{3g_{\pi NP_{b}}^{2}(m_{N}+E_{N})}{%
4\pi m_{P_{b}}}|\vec{p}_{N}^{~\mathrm{c.m.}}|,
\end{eqnarray}

with

\begin{eqnarray}
|\vec{p}_{\Upsilon }^{~\mathrm{c.m.}}| &=&\frac{\lambda
(m_{P_{b}}^{2},m_{\Upsilon }^{2},m_{N}^{2})}{2m_{P_{b}}}, \\
|\vec{p}_{N}^{~\mathrm{c.m.}}| &=&\frac{\lambda (m_{P_{b}}^{2},m_{\pi
}^{2},m_{N}^{2})}{2m_{P_{b}}}, \\
E_{N} &=&\sqrt{|\vec{p}_{N}^{~\mathrm{c.m.}}|^{2}+m_{N}^{2}},
\end{eqnarray}%
where $\lambda $ is the K\"{a}llen function with $\lambda (x,y,z)\equiv
\sqrt{(x-y-z)^{2}-4yz}$, and $\mathcal{M}_{P_{b}\rightarrow \Upsilon N}$ is
the decay amplitudes. The $m_{P_{b}}$, $m_{\Upsilon }$, $m_{N}$, and $m_{\pi
} $ are the masses of $P_{b}$, $\Upsilon $, nucleon, and pion meson,
respectively.

For the electromagnetic (EM) coupling $eh$ related to the $\gamma NP_{b}$
vertex, its value can be obtained from the strong coupling constant $%
g_{P_{b}\Upsilon N}$ within the VMD mechanism \cite%
{Bauer:1977iq,Bauer:1975bv,Bauer:1975bw}. The EM coupling
constants $eh$ are related to the coupling constants $g_{P_{b}\Upsilon N}$ as%
\begin{eqnarray}
eh &=&g_{P_{b}\Upsilon N}\frac{e}{f_{\Upsilon }}\frac{2m_{N}}{%
(m_{P_{b}}^{2}-m_{N}^{2})m_{\Upsilon }}  \notag \\
&&\times \sqrt{m_{\Upsilon
}^{2}(m_{N}^{2}+4m_{N}m_{P_{b}}+m_{P_{b}}^{2})+(m_{P_{b}}^{2}-m_{N}^{2})^{2}}%
,
\end{eqnarray}%
\ \ \

The Lagrangian depicting the coupling of the meson $\Upsilon $ with a photon
reads as%
\begin{equation}
\mathcal{L}_{\Upsilon \gamma }=-\frac{em_{\Upsilon }^{2}}{f_{\Upsilon }}%
V_{\mu }A^{\mu },
\end{equation}%
where $f_{\Upsilon }$ is the $\Upsilon $ decay constant, respectively. Thus
one gets the expression for the $\Upsilon \rightarrow e^{+}e^{-}$ decay,%
\begin{equation}
\Gamma _{\Upsilon \rightarrow e^{+}e^{-}}=\left( \frac{e}{f_{\Upsilon }}%
\right) ^{2}\frac{8\alpha \left\vert \vec{p}_{e}^{~\mathrm{c.m.}}\right\vert
^{3}}{3m_{\Upsilon }^{2}},
\end{equation}%
where $\vec{p}_{e}^{~\mathrm{c.m.}}$ denotes the three-momentum of an
electron in the rest frame of the $\Upsilon $ meson. The $\alpha
=e^{2}/4\pi=1/137$ is the electromagnetic fine structure constant. With
the partial decay width of $\Upsilon \rightarrow e^{+}e^{-}$ \cite%
{Tanabashi:2018oca}, $\Gamma _{\Upsilon \rightarrow e^{+}e^{-}}\simeq 1.34\text{ keV}$,
one gets $e/f_{\Upsilon }\simeq 0.0076$.

Since the partial decay widths $\Gamma _{P_{b}\rightarrow \Upsilon N}$ of
the $P_{b}$ states have been predicted in Ref.~\cite{Gutsche:2019mkg}, the
EM couplings related to the $\gamma NP_{b}$ vertices is also determined. In our previous work \cite{Wang:2019dsi}, the result shows that the
experimental data point near the threshold is consistent with the
contribution from the $P_{c}(4312)$ state by taking branching ratio
Br$[P_{c}\rightarrow \pi N]\simeq 0.05\%$. In this paper, a small
branching ratio of Br$[P_{b}\rightarrow \pi N]\simeq 0.05\%$ will
also be used to calculate the coupling constant $g_{_{\pi NP_{b}}}$. We will
discuss this branching ratio in detail in the next section in conjunction with
the values of cross section of $P_{b}$ in the reaction $\pi ^{-}p\rightarrow \Upsilon n$. The obtained coupling constants are listed in Table~\ref{tab1} by assuming that the $\pi N$ channel accounts for $0.05\%$ of total widths of $%
P_{b} $ states. \renewcommand\tabcolsep{0.26cm} \renewcommand{%
\arraystretch}{2}
\begin{table}[tbph]
\caption{The values of coupling constants by taking the decay width of $%
\Gamma _{P_{b}(11080)\rightarrow \Upsilon N}=0.38$ MeV and $\Gamma
_{P_{b}(11125)\rightarrow \Upsilon N}=3.27$ MeV, while assuming the $\protect\pi %
N$ channel account for $0.05\%$ of total widths of $P_{b}$ states.}
\label{tab1}%
\begin{tabular}{ccccc}
\hline\hline
states & $\Gamma _{P_{b}\rightarrow \Upsilon N}$ (MeV) & $g_{P_{b}\Upsilon
N} $ & $eh$ & $g_{P_{b}\pi N}$ \\ \hline
$P_{b}(11080)$ & 0.38 & 0.074 & 0.00016 & 0.00049 \\
$P_{b}(11125)$ & 3.27 & 0.213 & 0.00045 & 0.00145 \\ \hline\hline
\end{tabular}%
\end{table}

\subsection{Pomeron exchange and Reggeized $t$ channel}

\begin{figure}[h]
\begin{center}
\includegraphics[scale=0.41]{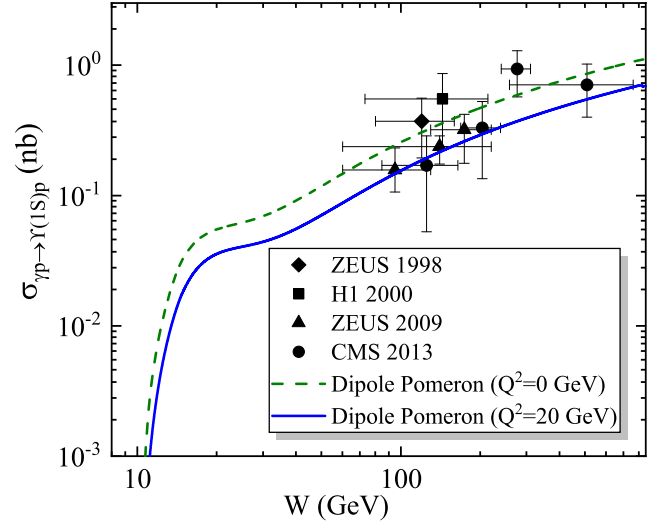}
\end{center}
\caption{Cross section for the reaction $\protect\gamma p\rightarrow
\Upsilon p$  with the dipole Pomeron model by taking the coefficient
$N_{\Upsilon }=\protect\sqrt{2}/3$. The green dashed, blue solid lines are
for the Pomeron contributions with the $Q^{2}=0$ and 20 GeV, respectively.
The experimental data are from Refs.~\protect\cite%
{Breitweg:1998ki,Chekanov:2009zz,Adloff:2000vm,CMS:2016nct}.}
\label{dipole Pomeron}
\end{figure}

For the $P_{b}$ state photoproduction process, the Pomeron exchange is
considered as the main background contribution. In this work, the dipole
Pomeron model~\cite{Martynov:2001tn,Martynov:2002ez,Cao:2019gqo} is employed to
calculate the cross section from the background contribution. The basic
diagram is depicted in Fig. 1(c), where the $Q^{2}$ is the virtuality of the
photon, and $s$ and $t$ are the Mandelstam variables. Generally, since the
Regge model plays an important role in the high-energy process, a double
Regge pole is also included in the dipole Pomeron model. Using the dipole
Pomeron model and related parameters~\cite{Martynov:2002ez}, the cross
section of the reaction $\gamma p\rightarrow \Upsilon p$  can be calculated.
Moreover, in Ref.~\cite{Martynov:2002ez}, the result suggests that it may be
more appropriate when the value of the coefficient $N_{\Upsilon }$ is
between $N_{\phi }$ and $N_{J/\psi }$. In our calculation, by taking $%
N_{\Upsilon }=\sqrt{2}/3$, the obtained cross section in dipole Pomeron
model is well consistent with the experimental data~\cite%
{Breitweg:1998ki,Chekanov:2009zz,Adloff:2000vm,CMS:2016nct}, as shown in
Fig.~\ref{dipole Pomeron}.

For the $P_{b}$ production via the reaction $\pi ^{-}p\rightarrow \Upsilon n$,
the  background is mainly come from the $t$ channel $\pi
$, $\rho $, and $\omega $ exchanges, as depicted in Fig.~\ref{Fig: Feynman2}(c). Since the
branching ratios of $\Upsilon $ decaying into $\rho \pi $ and $\omega \pi $ channels are of the order of magnitude of $10^{-6}$, the contributions from the $\rho $ and $\omega $ exchanges
should be very small and can be ignored. Therefore, in our calculation,
only the contribution from the $\pi $ exchange is considered. The
effective Lagrangians for the vertices of $\Upsilon \pi \pi $ and $\pi NN$
read as
\begin{eqnarray}
\mathcal{L}_{\Upsilon \pi \pi } &=&-ig_{_{\Upsilon \pi \pi }}(\pi
^{-}\partial _{\mu }\pi ^{+}-\partial _{\mu }\pi ^{-}\pi ^{+})\Upsilon ^{\mu
}, \\
\mathcal{L}_{\pi NN} &=&-ig_{_{\pi NN}}\bar{N}\gamma _{5}\vec{\tau}\cdot
\vec{\pi}N,
\end{eqnarray}%
where ${N}$, $\pi $ and $\Upsilon $ are the nucleon, the pion, and the $%
\Upsilon $ meson fields, respectively. Here, the $g_{_{\pi NN}}^{2}/4\pi
=12.96$ is adopted \cite{Lin:1999ve}. Moreover, the coupling constant $%
g_{_{\Upsilon \pi \pi }}$ can be determined from the decay width%
\begin{equation*}
\Gamma _{\Upsilon \rightarrow \pi \pi }=\frac{g_{_{\Upsilon \pi \pi }}^{2}}{%
6\pi m_{\Upsilon }^{2}}|\vec{q}|^{3},
\end{equation*}%
where $|\vec{q}|=\sqrt{m_{\Upsilon }^{2}-4m_{\pi }^{2}}/2$. Taking the value
of decay width $\Gamma _{\Upsilon \rightarrow \pi \pi }$ as $0.027$ keV \cite%
{Tanabashi:2018oca}, one obtain the coupling constant $g_{_{\Upsilon \pi \pi
}}\simeq 0.0007$.

Since the energy corresponding to the $\pi ^{-}p\rightarrow \Upsilon n$
reaction is above 10 GeV, the Reggeized treatment will be applied to the $t$
channel process. Usually, one just need to replace the Feynman propagator
with the Regge propagator as%
\begin{equation}
\frac{1}{t-m_{\pi }^{2}}\rightarrow (\frac{s}{s_{scale}})^{\alpha _{\pi }(t)}%
\frac{\pi \alpha _{\pi }^{\prime }}{\Gamma \lbrack 1+\alpha _{\pi }(t)]\sin
[\pi \alpha _{\pi }(t)]},
\end{equation}%
where the scale factor $s_{scale}$ is fixed at 1 GeV. In addition, the Regge
trajectories of $\alpha _{\pi }(t)$ is written as \cite{Wang:2019dsi},%
\begin{equation}
\alpha _{\pi }(t)=0.7(t-m_{\pi }^{2}).\quad \ \
\end{equation}%
It can be seen that no free parameters have been added after the
introduction of the Regge model.

\subsection{Amplitude}

Based on the Lagrangians above, the scattering amplitude for the reactions $\gamma
p\rightarrow \Upsilon p$ and $\pi ^{-}p\rightarrow \Upsilon n$  can
be constructed as%
\begin{eqnarray}
-i\mathcal{M}_{\gamma p\rightarrow \Upsilon p} &=&\epsilon _{\Upsilon }^{\nu
}(k_{2})\bar{u}(p_{2})\mathcal{A}_{\mu \nu }u(p_{1})\epsilon _{\gamma }^{\mu
}(k_{1}), \\
-i\mathcal{M}_{\pi ^{-}p\rightarrow \Upsilon n} &=&\epsilon _{\Upsilon
}^{\mu }(k_{2})\bar{u}(p_{2})\mathcal{B}_{\mu }u(p_{1}),
\end{eqnarray}%
where $u$ is the Dirac spinor of nucleon, and $\epsilon _{\Upsilon }$ and $%
\epsilon _{\gamma }$ are the polarization vector of $\Upsilon $ meson and
photon, respectively.

The reduced amplitude $\mathcal{A}_{\mu \nu }$ for the $s$ channel $P_{b}$
photoproduction reads
\begin{equation}
\mathcal{A}_{\mu \nu }^{s} =\frac{eh}{2m_{N}}g_{P_{b}\Upsilon N}\mathcal{F}%
_{s}(q_{s}^{2})\gamma _{5}\gamma _{\nu }\frac{(\rlap{$\slash$}q_{s}+m_{P_{b}})}{%
q_{s}^{2}-m_{P_{b}}^{2}+im_{P_{c}}\Gamma _{P_{b}}}  \gamma _{\mu }\rlap{$\slash$}k_{1},\label{AmpT1}
\end{equation}
and  the reduced amplitude $\mathcal{B}_{\mu }$ for the $s$ channel with $%
P_{b}$ and $t$ channel with $\pi $ exchanges are written as
\begin{eqnarray}
\mathcal{B}_{\mu }^{s} &=&\sqrt{2}g_{_{\pi NP_{b}}}g_{P_{b}\Upsilon N}%
\mathcal{F}_{s}(q_{s}^{2})\gamma _{5}\gamma _{\mu }\frac{(\rlap{$\slash$}%
q_{s}+m_{P_{b}})}{q_{s}^{2}-m_{P_{b}}^{2}+im_{P_{c}}\Gamma _{P_{b}}}, \\
\mathcal{B}_{\mu }^{t} &=&\sqrt{2}g_{_{\Upsilon \pi \pi }}g_{_{\pi NN}}\frac{%
\mathcal{F}_{t}(q_{t}^{2})}{q_{t}^2-m_{\pi }^{2}}\gamma _{5}k_{1\mu },
\end{eqnarray}
where  $q_{s}=k_{1}+p_{1}$ and $q_{t}=k_{1}-k_{2}$  are the four-momenta of the exchanged $P_{b}$ state in $s$ channel and $\pi$ meson in $t$ channel, respectively.

For the $s$ channel with intermediate $P_{b}$ state, one adopts a general
form factor to describe the size of hadrons as \cite%
{Kim:2011rm,Wang:2019krd,Wang:2019dsi},
\begin{equation}
\mathcal{F}_{s}(q_{s}^{2})=\frac{\Lambda ^{4}}{\Lambda
^{4}+(q_{s}^{2}-m_{P_{b}}^{2})^{2}}~,
\end{equation}%
where $q_{s}$ and $m_{P_{b}}$ are the four-momentum and mass of the
exchanged $P_{b}$ state, respectively. For the heavier hadron production,
the typical value of cutoff  $\Lambda =0.5$ GeV will be{\ taken }as used in
Refs. \cite{Kim:2011rm,Wang:2019krd,Wang:2019dsi}.

For the $t$-channel meson exchanges \cite{Kim:2011rm,Wang:2019dsi}, the
general form factor $\mathcal{F}_{t}(q_{t}^{2})$ consisting of $\mathcal{F}%
_{\Upsilon \pi \pi }=(\Lambda _{t}^{2}-m_{\pi }^{2})/(\Lambda
_{t}^{2}-q_{\pi }^{2})$ and $\mathcal{F}_{\pi NN}=(\Lambda _{t}^{2}-m_{\pi
}^{2})/(\Lambda _{t}^{2}-q_{\pi }^{2})$ are taken into account. Here, $%
q_{\pi }$ and $m_{\pi }$ are 4-momentum and mass of the $\pi $ meson,
respectively. The value of the cutoff $\Lambda _{t}$ is taken as 2.0 GeV,
which is the same as that in Ref. \cite{Wang:2019dsi}.

\section{Numerical results}

With the preparation in the previous section, the cross section of the reaction $%
\gamma p\rightarrow \Upsilon p$  can be calculated. The differential
cross section in the center of mass (c.m.) frame is written as
\begin{equation}
\frac{d\sigma }{d\cos \theta }=\frac{1}{32\pi s}\frac{\left\vert \vec{k}%
_{2}^{{~\mathrm{c.m.}}}\right\vert }{\left\vert \vec{k}_{1}^{{~\mathrm{c.m.}}%
}\right\vert }\left( \frac{1}{J}\sum\limits_{\lambda }\left\vert \mathcal{M}%
\right\vert ^{2}\right) ,
\end{equation}%
with $J=4$ for the $\gamma p\rightarrow \Upsilon p$ reaction and $J=2$ for
the reaction $\pi ^{-}p\rightarrow \Upsilon n$. Here, $s=(k_{1}+p_{1})^{2}$,
and $\theta $ denotes the angle of the outgoing $\Upsilon $ meson relative
to $\pi /\gamma $ beam direction in the c.m. frame. $\vec{k}_{1}^{{~\mathrm{%
c.m.}}}$ and $\vec{k}_{2}^{{~\mathrm{c.m.}}}$ are the three-momenta of the
initial photon beam and final $\Upsilon $ meson, respectively.

\subsection{$P_{b}$ production in reaction $\protect\gamma p\rightarrow \Upsilon p$}

\begin{figure}[h]
\begin{center}
\includegraphics[scale=0.4]{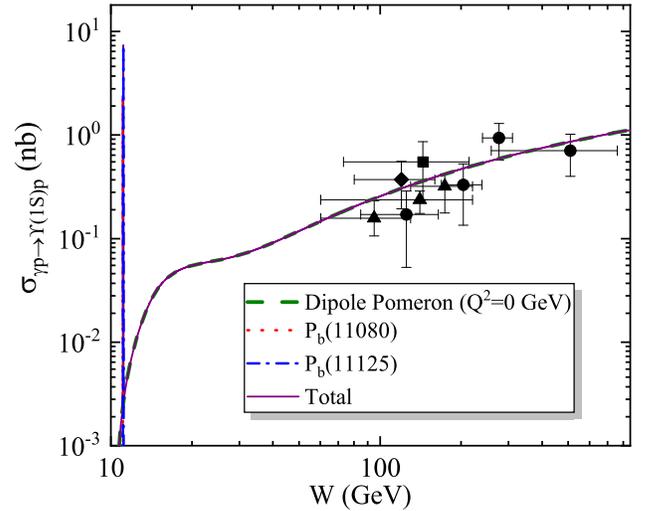}
\end{center}
\caption{Total cross section for the reaction $\protect\gamma p\rightarrow \Upsilon p$.
The green dashed, red dotted, blue dot-dashed, and violet solid lines are for
the Pomeron, the $P_{b}(11080)$, the $P_{b}(11125)$, and total contributions,
respectively. The experimental data are taken from Refs.~\protect\cite%
{Breitweg:1998ki,Chekanov:2009zz,Adloff:2000vm,CMS:2016nct}.}
\label{total01}
\end{figure}

In Fig.~\ref{total01} we present the total cross section for the reaction $\gamma
p\rightarrow \Upsilon p$ from threshold to 800 GeV of the center of
mass energy. It is found that the contribution from pentaquark $P_{b}$ state
shows a very sharp peak near the threshold, and the experimental point at
high energy is well consistent with the cross section of the Pomeron
diffractive process.

\begin{figure}[h]
\begin{center}
\includegraphics[scale=0.4]{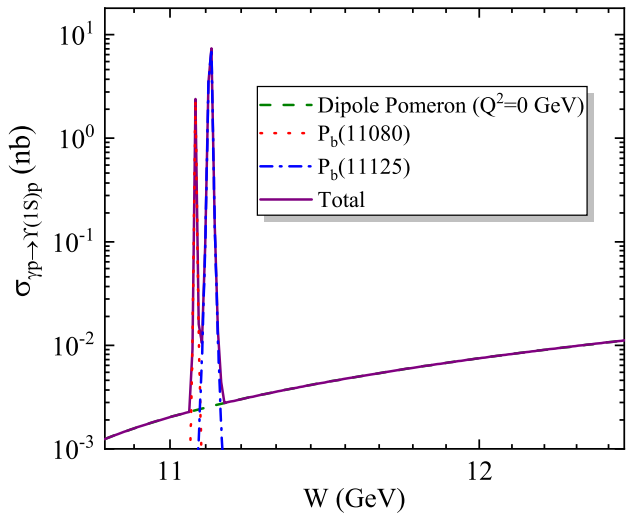}
\end{center}
\caption{Same as Fig. 4 except that the energy range is
reduced.}
\label{total02}
\end{figure}

\begin{figure}[h]
\begin{center}
\includegraphics[scale=0.4]{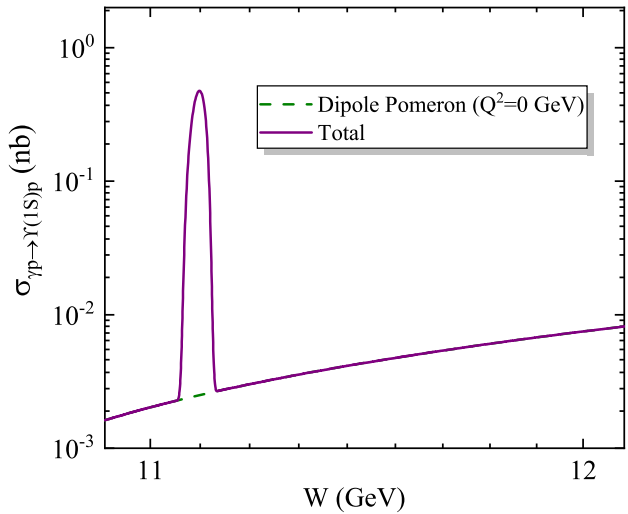}
\end{center}
\caption{Same as Fig. 5 except that the energy bin width is
increased to 0.1 GeV.}
\label{total03}
\end{figure}

In order to more clearly distinguish the contributions from the $P_{b}$
states, in Fig.~\ref{total02} we give the same results as Fig.~\ref{total01}
but with a reduced energy range. As can be seen from Fig.~\ref{total02},
when the energy bin width is chosen as 0.01 GeV, two distinct peaks can be seen around
11 GeV, which are derived from the contributions of  states $P_{b}(11080)$ and $%
P_{b}(11125)$, respectively. Since the mass of $P_{b}(11080)$ and $%
P_{b}(11125)$ are very close to each other, if we increase the bin width to 0.1 GeV, it is difficult to distinguish the two peaks, and a larger bump will be formed, as shown in Fig.~\ref{total03}. Therefore, if different $P_{b}$ signals need to be distinguished, a small energy bin width is needed.

Moreover, the result suggests that the theoretical average of the cross section
from the  $P_{b}(11080)$ is of a order of  magnitude of 1 nb if the bin width is 0.01 GeV. If the bin width increases to 0.1 GeV, the theoretical average value
of the cross section of $P_{b}(11080)$  becomes about 0.3 nb. In  Refs.~\cite{Karliner:2015voa,Aid:1996dn},
the cross section of  process $\gamma p\rightarrow
\Upsilon p$ at 11 GeV was estimated to be about 12 pb $-$ 50 pb. Even if the background cross
section can reach this magnitude, as long as the bin width is less than or equal to 0.1
GeV,  the possibility of distinguishing the $P_{b}$ state from the
background is great.

\subsection{$P_{b}$ production in reaction $\protect\pi ^{-}p\rightarrow \Upsilon n$
}

\begin{figure}[h]
\begin{center}
\includegraphics[scale=0.4]{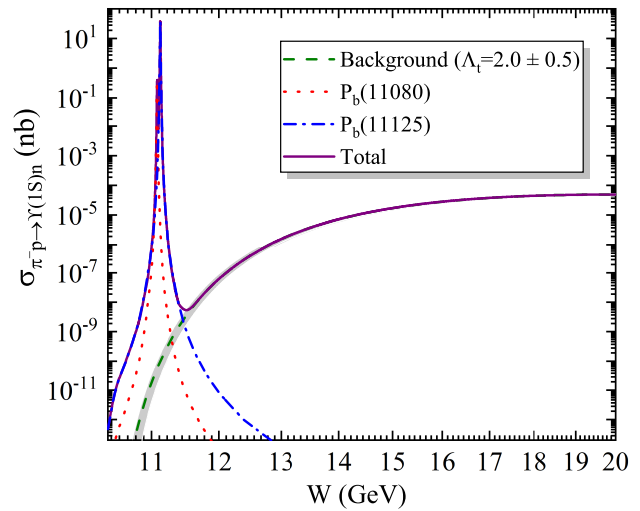}
\end{center}
\caption{Total cross section for the reaction $\protect\pi ^{-}p\rightarrow \Upsilon n$.
The green dashed, red dotted, blue dot-dashed, and violet solid lines are for
the background, the $P_{b}(11080)$, the $P_{b}(11125)$, and total contributions,
respectively. The bands stand for the error bar of the cutoff $\Lambda_{t}$.}
\label{total04}
\end{figure}

In Fig.~\ref{total04}, the total cross section of the interaction $\pi ^{-}p\rightarrow
\Upsilon n$ is presented. It can be seen that near the threshold,
the background cross section from the $t$ channel is very small, even if we
take a value of cutoff as  2.0 GeV. The main reason for
this situation is that the branching ratio of $\Upsilon $ decaying into $\pi \pi $ channel
is very small, and the branching ratios of $\Upsilon $ decaying to $\rho \pi $
and $\omega \pi $ channels are smaller and can be ignored. Near the threshold, we can
see very sharp peaks, which come from the contribution of the $P_{b}$ states.

\begin{figure}[h]
\begin{center}
\includegraphics[scale=0.4]{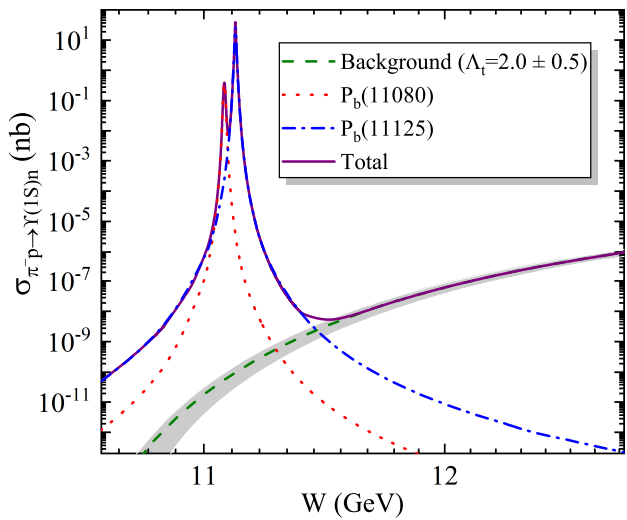}
\end{center}
\caption{Same as Fig. 7 except that the energy range is
reduced.}
\label{total05}
\end{figure}

To more clearly distinguish the cross section at the threshold, we present more explicit near the threshold in
 Fig.~\ref{total05}, which is the same as the Fig.~\ref{total04} except
that the energy range is reduced. One find that the total cross section
exhibits two peaks around the 11 GeV when the bin width is taken to be 0.01
GeV. According to our calculation, when the bin width is 0.1 GeV, branching
ratio of  Br$[P_{b}\rightarrow \pi N]\simeq 0.05\%$, the
theoretical average value of the cross section from the $P_{b}(11080)$ state
reaches at least 0.1 nb in a bin interval, which is several orders of
magnitude higher than the background cross section.

\begin{figure}[h]
\begin{center}
\includegraphics[scale=0.4]{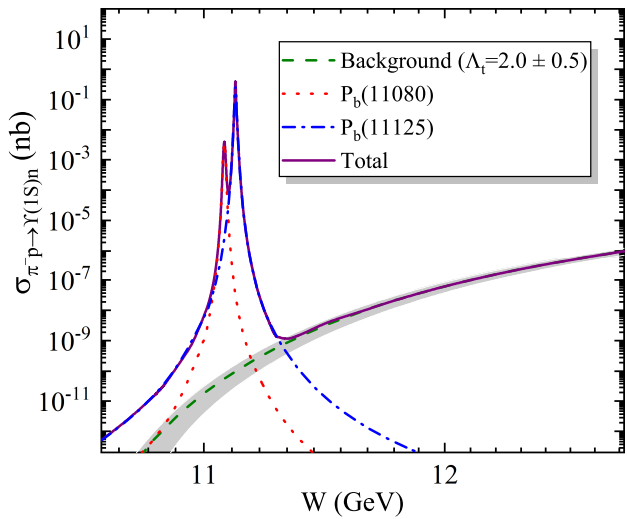}
\end{center}
\caption{Same as Fig. 8 except that the branching ratio of $P_{b}$ to $\pi N$ is
reduced to $0.0005\%$.}
\label{total06}
\end{figure}

Specifically, when we take the branching ratio of $P_{b}$ to $\pi N$ as $0.0005\%$, the corresponding coupling constants $g_{P_{b}(11080)\pi N}$ and $g_{P_{b}(11125)\pi N}$ are 0.000049 and 0.000145, respectively.  According to the obtained coupling constants, the total cross section of the reaction $\pi ^{-}p\rightarrow \Upsilon n$  is given when the energy bin width is 0.01 GeV, as shown in Fig. ~\ref{total06}. One notice that the signal is still much higher than the background. Even increasing the energy bin width to 0.1 GeV, the theoretical average of the cross section from the $P_{b}(11080)$ can reach the order of 0.001 nb,
which means that the signal is still clearly distinguished from the background.
\section{Summary and discussion}

The pentaquark is an important kind of the exotic hadron. The observation of the $P_c$ states at LHCb inspires us to predict the possible hidden-bottom pentaquark.  Since it is impossible to decay from a heavy hadrons as $P_c$ states,  it is interesting to study production of hidden-bottom pentaquarks through exciting  nucleon by photon or pion.  In this work, based on the effective field theory and the VMD
mechanism, the production of the hidden-bottom pentaquark $P_{b}$ states via $%
\gamma p$ or $\pi ^{-}p$ scattering is investigated.

For the background of  the $P_{b}$ state photoproduction, it is thought that it
mainly comes from the contribution of the Pomeron exchange. Using the dipole
pomeron model, the experimental data of the interaction $\gamma p\rightarrow \Upsilon p$
 can be well described. Near the threshold, when the bin width is
0.01 GeV,  very sharp peaks from the $P_{b}$
states can be clearly distinguished  in the cross sections. According to our calculation, even if the
bin width is increased to 0.1 GeV, the cross section of the $P_{b}$ signal
is still higher than the background, which means that the the vicinity of
the center of mass energy of 11 GeV is the best energy window for searching
for these $P_{b}$ states via the photoproduction process. It is suggested that
experiments to search for the $P_{b}$ states through the $\gamma p$ scattering can be
carried out, which is well within the capabilities of the COMPASS facility
at CERN \cite{Nerling:2012er}.

For the production of the $P_{b}$ states in the reaction $\pi ^{-}p\rightarrow \Upsilon n$, the background is considered to be mainly derived from the
contribution of the Reggeized $t$ channel with $\pi $ exchange. Since the
branching ratio of the $P_{b}$ decaying into $\pi \pi $ channel is very small, the
background cross section of the process $\pi ^{-}p\rightarrow \Upsilon n$  is
several orders of magnitude lower than the cross section from the $P_{b}$
signal. According to our calculation, even if the branching ratio of the
 the $P_{b}$ into $\pi N$ channel is reduced to $0.0005\%$, the average value of the
cross section from $P_{b}(11080)$ contribution reaches an order of 0.001
nb/100 MeV, which is still much higher than the background contribution.
Compared to the photoproduction process, the requirement of experimental
precision for searching for these $P_{b}$ states by the $\pi ^{-}p$ scattering
may be lower, which indicates that searching for the process $P_{b}$ via $\pi
^{-}p\rightarrow \Upsilon n$  may be a very important and promising
way. Hence, an experimental study of the hidden bottom pentaquark $P_{b}$
states via the pion-induced reaction is strongly suggested, which can be
carried on the J-PARC and COMPASS facilites \cite{Nerling:2012er,Kumano:2015gna}.

In addition, since both the electromagnetic $\gamma p$ and $ep$ reactions
are very similar, we also look forward to finding and studying these hidden
bottom pentaquark $P_{b}$ states through the $ep$ scattering process.
Electromagnetic probes covering the above energy are expected to be produced
on future EicC (Electron Ion Collider in China) facility in China. The accumulated luminosity with one-year run is about 50fb$^{-1}$~\cite{Chen:2018wyz,Chen:2019equ}. Because such facility is usually impossible to run near a fixed energy point for a year, we make an estimation of one-day run, that is, an accumulated luminosity about  0.1~fb$^{-1}$. The number of events  reaches  $1\times10^{5}$/0.01 GeV near the mass of the $P_b$ state  if we adopt a cross section 1nb/0.01 GeV predicted in our calculation. Even after considering the efficiency and the uncertainty of theoretical prediction, which may depress the number of event by two or three orders of magnitude,  the observation of  hidden-bottom pentaquarks is still very promising at EicC.

\section{Acknowledgments}
This project is supported by the National Natural Science Foundation of
China under Grants No. 11705076 and No. 11675228.
We acknowledge the supported by the Key Research Program of the Chinese Academy of Sciences, Grant NO. XDPB09.
This work is partly supported by HongLiu Support Funds for Excellent Youth Talents of Lanzhou University of Technology.

\end{document}